\documentclass[prb,preprint,aps,superscriptaddress,longbibliography]{revtex4-1}

\usepackage{graphicx, epstopdf}
\usepackage{subfigure}
\usepackage{float}
\usepackage{amsmath}
\usepackage{amsfonts}
\usepackage{amssymb}
\usepackage{bm}% bold math
\usepackage{subfigure}
\usepackage{caption}
\usepackage{hyperref}
\begin{document}

\title{Topological Semimetals with Triply Degenerate Nodal Points in $\theta$-phase Tantalum Nitride}

%\footnote{These authors contribute equally to this paper.}}

\author{Hongming Weng}
\email{hmweng@iphy.ac.cn}
\affiliation{Beijing National Laboratory for Condensed Matter Physics,
  and Institute of Physics, Chinese Academy of Sciences, Beijing
  100190, China}
\affiliation{Collaborative Innovation Center of Quantum Matter,
  Beijing, China}

\author{Chen Fang}
\email{cfang@iphy.ac.cn}
\affiliation{Beijing National Laboratory for Condensed Matter Physics,
  and Institute of Physics, Chinese Academy of Sciences, Beijing
  100190, China}

\author{Zhong Fang}
%\email{zfang@iphy.ac.cn}
\affiliation{Beijing National Laboratory for Condensed Matter Physics,
  and Institute of Physics, Chinese Academy of Sciences, Beijing
  100190, China}

\affiliation{Collaborative Innovation Center of Quantum Matter,
  Beijing, China}

\author{Xi Dai}
%\email{daix@iphy.ac.cn}

\affiliation{Beijing National Laboratory for Condensed Matter Physics,
  and Institute of Physics, Chinese Academy of Sciences, Beijing
  100190, China}

\affiliation{Collaborative Innovation Center of Quantum Matter, Beijing, China}

\date{\today}

\begin{abstract}
Using first-principles calculation and symmetry analysis, we propose that $\theta$-TaN is a topological semimetal having a new type of point nodes, i.e., triply degenerate nodal points. Each node is a band crossing between degenerate and non-degenerate bands along the high-symmetry line in the Brillouin zone, and is protected by crystalline symmetries. Such new type of nodes will always generate singular touching points between different Fermi surfaces and 3D spin texture around them. Breaking the crystalline symmetry by external magnetic field or strain leads to various of topological phases. By studying the Landau levels under a small field along $c$-axis, we demonstrate that the system has a new quantum anomaly that we call ``helical anomaly''.
\end{abstract}

%\pacs{71.20.-b, 73.20.-r, 73.43.-t}
\maketitle

\section{Introduction} \label{introduction}
The discovery of topological semimetals (TSM) is one of the major progress in condensed matter physics within the last decade.~\cite{TSM_review,Chiu_RMP,Bansil_RMP}
 The type of a TSM is determined according to the symmetry that protects the band crossing point near the Fermi energy and the effective Hamiltonian near that point. 
 For example, Dirac semimetal~\cite{young_dirac_2012, Na3Bi, Cd3As2} is characterized by two bands with double degeneracy that cross 
 near the Fermi level, and has to be protected by certain crystalline symmetry either at high symmetry point or along high symmetry lines.~\cite{yang_classification_2014} 
 In contrast, the formation of Weyl semimetal,~\cite{murakami_phase_2007,wan,HgCrSe,TaAs_Weng,HuangSM_Weyl,Lu622,TaAs_arc, TaAs_node, Xu2015a, Yang2015,Weyl_arc_ossi,Soluyanov2015} 
 which is characterized by the crossing of two non-degenerate bands at the Fermi level, does not require any protection from the crystalline 
 symmetry other than lattice translation. In fact, the Weyl points in Weyl semimetals can be viewed as the ``topological defects" in momentum 
 space, which are stable under continuous deformation of the Hamiltonian.~\cite{fang_anomalous_2003, adv_phys} Besides Dirac and Weyl 
 semimetals, nodal line semimetal is another type of TSM where two bands cross each other along a line in the BZ.~\cite{burkov,allcarbon_nodeLine2014, LnX,Cu3NPd,Cu3NPdKane,CFang_NLSM_PRB}

Besides the above mentioned Weyl, Dirac and nodal line semimetals, there are other types of TSM protected by nonsymmorphic space group 
symmetries, which are characterized by three-, six- or eight-fold degenerate points at the Fermi level and named as ``new fermions" by 
Bradlyn {\it et al.}~\cite{newfermion} In the present paper, a new mechanism to generate ``new fermions" is proposed with a realistic 
material $\theta$-TaN in WC-type structure. In the band structure of $\theta$-TaN~\cite{thetaTaN-2} or similar materials,~\cite{NbN_1989,NbN_1990} 
along a certain high symmetry axis, both one and two dimensional representations are allowed, which makes it possible to generate band 
crossing between a doubly degenerate band and a non-degenerate band near the Fermi level at a triply degenerate nodal point (TDNP). This new type of three-component 
fermions can be viewed as the ``intermediate species'' between the four-component Dirac and the two-component Weyl fermions. 

From another point of view, all the above listed TSM can also be characterized by the topological features of the Fermi surface (FS) with the 
Fermi level near the band crossing points. For example, in Weyl semimetals the FS is non-degenerate with a nonzero Chern number,~\cite{HgCrSe,adv_phys} 
while in Dirac semimetals the FS is doubly degenerate and can be viewed as two FS with opposite Chern numbers located on top of 
each other~\cite{Na3Bi,adv_phys}. Compared to Weyl and Dirac semimetals, the FS in $\theta$-TaN can be characterized by two non-degenerate 
FS touching at one single point. Unlike the situation in the type-II Weyl semimetal state~\cite{Soluyanov2015}, where the FS touching 
appears only when the Fermi level is right at the Weyl point, in $\theta$-TaN, the FS touching happens for a large range of chemical potential. 
Moreover a unique pattern of spin-momentum locking is found on the 3D FS in $\theta$-TaN, required by the crystalline symmetries.

For Weyl semimetals, the emergent ``chiral anomaly'' is related characteristic transport properties under external magnetic field, i.e. the negative 
magneto-resistance along the direction of the magnetic field.~\cite{Son2013,TaAs_anomaly, Transport_Weyl_XLQi_2013} In the quantum mechanical treatment for a single Weyl point under magnetic field, the chiral anomaly manifests itself in the presence of a chiral zeroth Landau level propagating along the direction of the field. In the present paper, we show that the Landau levels in $\theta$-TaN exhibits a ``helical anomaly'', manifested by the presence of a pair of counter-propagating modes under an external field along the high-symmetry direction, the crossing of which is protected by the threefold rotation symmetry.

\section{Computational Details} \label{method}
We have employed the software package OpenMX~\cite{openmx} for most of the first-principles calculations.
Exchange-correlation potential is treated within the generalized gradient approximation (GGA) of Perdew-Burke-Ernzerhof type.\cite{Perdew1996}
Spin-orbit coupling (SOC) is taken into account self-consistently. The sampling of the Brillouin zone in the self-consistent process is taken as the grid of
12$\times$12$\times$10. The basis set for Ta and N is chosen as Ta9.0-s2p2d2f1 and N7.0-s2p2d1, respectively. The crystal structure and the stability 
of $\theta$-TaN~\cite{thetaTaN-2} has been recently revisited by Friedrich {\it et al.}~\cite{thetaTaN}. The experimental crystal structure 
is fully relaxed until the residual forces on each atom is less than 0.001 eV/\AA. The possible underestimation of band gap within GGA 
is checked by non-local Heyd-Scuseria-Ernzerhof (HSE06) hybrid functional\cite{heyd2003hybrid, heyd2006hybrid} calculation using VASP software 
package.~\cite{kresse1996_1, kresse1996_2} To explore the surface states, we construct the maximally localized Wannier functions (MLWF)~\cite{marzari1997,souza2001} 
for $d$ orbitals of Ta by using OpenMX.~\cite{openmx, weng_mlwf} They are used as basis set to build a tight-binding model for the the semi-infinite system 
with surface in Green's function method.~\cite{MRS_weng:9383312, adv_phys} 

\begin{figure}
\includegraphics[width=0.8\textwidth]{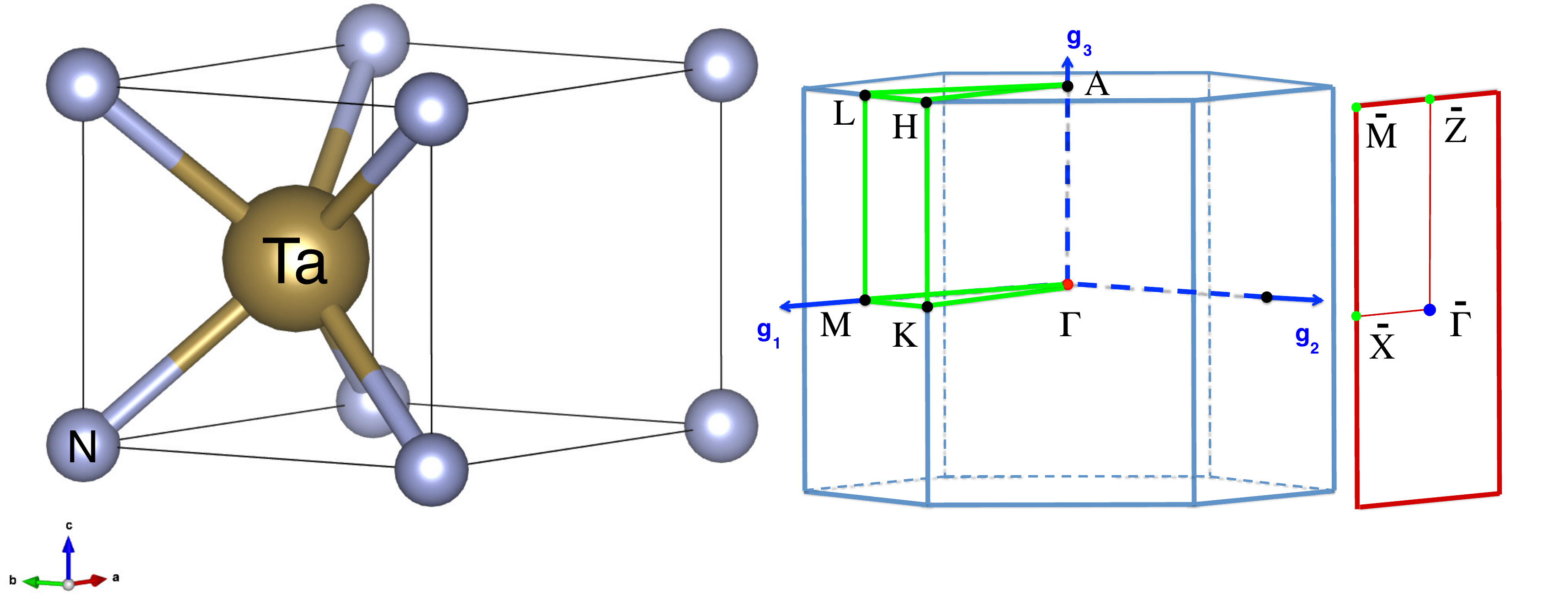}
\caption{(Color online)  (a) Crystal structure of $\theta$-TaN. (b) 3D bulk Brillouin zone (BZ) and projected (100) surface BZ with high symmetry crystal momenta indicated. }
\label{crystructure}
\end{figure}

\section{Results and Discussion} \label{Results}
% the crystal structure
{\it Crystal Structure.} 

The elements Ta and N can form many tantalum nitride phases.~\cite{thetaTaN-2, thetaTaN} $\theta$-TaN can be synthesized at high 
pressure (2-10 GPa) within a proper high temperature range. After cooling and pressure relaxation, it can be stabilized and shows WC-type hexagonal 
crystal structure with space group $P\bar{6}m2$ (No. 187). Ta and N are at the 1$d$ (1/3, 2/3, 1/2) and 1$a$ (0,0,0) Wyckoff position, respectively. 
The experimental lattice constants are $a$=$b$=2.9333(1) \AA~and $c$=2.8844(2) \AA.~\cite{thetaTaN} The theoretical relaxed lattice 
constants are $a$=$b$=2.9697~\AA~and $c$=2.9190 \AA, which are both overestimated by about 1.2\% and used in the following 
calculations. NbN can also be crystalized in the same WC-type structure.~\cite{NbN_1989,NbN_1990} 

{\it Band structure of $\theta$-TaN.} 

\begin{figure}
\includegraphics[width=0.65\textwidth]{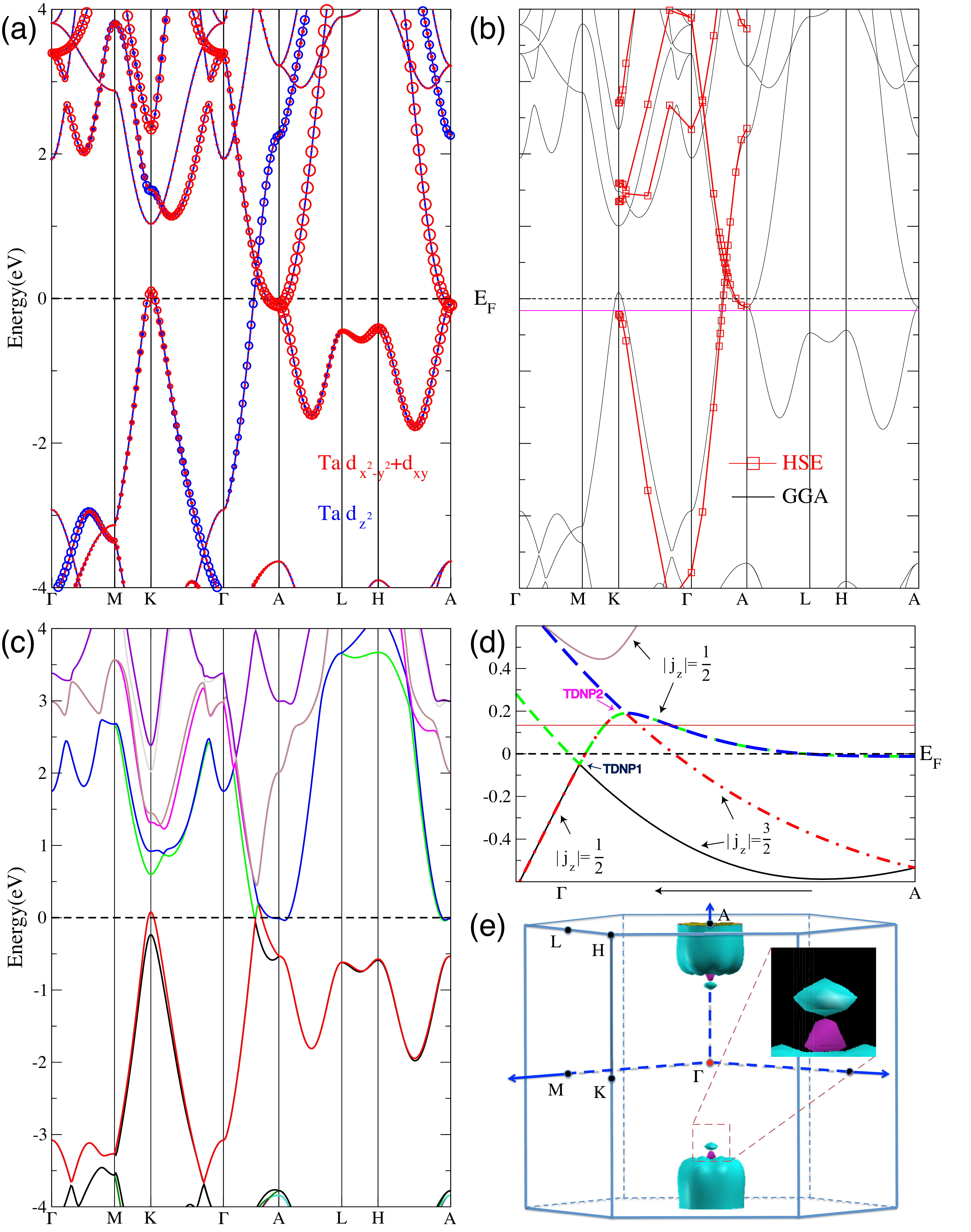}
\caption{(Color online) Band structure of $\theta$-TaN within GGA (a) with fatted bands projected on to Ta $d_{z^2}$ 
and Ta $e_g$ ($d_{x^2-y^2}$ and $d_{xy}$) orbitals and (b) in comparison with that calculated by using hybrid functional HSE06. (c) Band structure with SOC included. (d) Enlarged band structure along $\Gamma$-A in (c) around TDNP. (e) The Fermi surface with chemical potential at 110 meV within GGA+SOC.}
\label{TaN_band}
\end{figure}

Fig.~\ref{TaN_band}(a) shows that $\theta$-TaN is a semimetal with both hole and electron Fermi pockets. There is a band crossing 
along $\Gamma$-A. Without considering SOC, the fatted bands clearly show that the crossing bands are one non-degenerate band composed of Ta $d_{z^2}$ orbital 
and a double degenerate band from $e_g$ orbitals ($d_{xy}$ and $d_{x^2-y^2}$ orbitals). The crossing point is exactly threefold degenerate protected by 
the $C_3$ rotational symmetry on $\Gamma$-A. Such band crossing is due the band inversion between the $d_{z^2}$ state and 
the $e_g$ states at A, which is similar to the case in Dirac semimetal Na$_3$Bi and Cd$_3$As$_2$~\cite{Na3Bi, Cd3As2}. To overcome
the possible overestimation of the band inversion within GGA, the hybrid functional HSE06 is used. It is found that this band inversion
remains and furthermore, the hole pocket at momentum K within GGA disappears with the band maximum pushed down to 
lower than the Fermi level. Since Ta is heavy and the SOC cannot be ignored, we further calculate the band structure with SOC included
self-consistently. Due to the lack of inversion symmetry, the spin splitting of bands at general momenta can be seen in Fig. ~\ref{TaN_band}(c). With SOC considered, the $d_{z^2}$ orbital contains two states with $J_z=\pm1/2$, where $J_z$ is the total angular momentum. The two $e_g$ orbitals contains four states with $J_z=\pm1/2,\pm3/2$. Due to the crystalline symmetries, the four $|J_z|=1/2$ states form two doublets, while the two $J_z=\pm3/2$ states are nondegenerate. (Here we remark that $3/2$ and $-3/2$ are equivalent because $C_3$-symmetry only preserves $J_z$ up to a multiple of 3.) In the Brillouin zone, near A, these six states form six bands near the Fermi energy. Along $\Gamma$-A, due to the vertical mirror symmetry, the $|J_z|=1/2$ bands are doubly degenerate, while $|J_z|=3/2$ bands are non-degenerate; and due to $C_3$-symmetry, bands having different $J_z$ cannot hybridize with each other [see Fig.~\ref{TaN_band}(d)]. This leads to two protected triply degenerate nodal
points (TDNPs) along $\Gamma$-A. We have also drawn the Fermi surfaces containing these two TDNPs by setting the chemical
potential in between them around 110 meV. The two Fermi surfaces, in diamond and bell shapes, respectively, touch each other
due to the double degeneracy of the $|J_z|=1/2$ band. The bigger cylinder-like Fermi surface
centering A is trivial since it doesn't enclose any band crossing points.

Some general remarks on the TDNP are due. First, the TDNP appear in pairs due to the time-reversal symmetry. Second, for any Fermi level that is not far from the TDNP energy, the Fermi surface consists at least of two pockets touching at a point along $\Gamma$-A. Finally, while the TDNP itself is protected by $C_3$ and vertical mirror symmetry, a small perturbation breaking these crystalline symmetries cannot fully gap the system, because the Fermi surface has a finite size for any chemical potential. This is in contrast with Dirac semimetals protected by crystalline symmetries, where an infinitesimal symmetry breaking perturbation can open a full gap at the Dirac point. This is also in contrast with type-II Weyl semimetals, where the touching of the electron and the hole pockets only appears when the Fermi level equals the energy of the Weyl point.

{\it Band topology and surface states of $\theta$-TaN.} 

\begin{figure}
\includegraphics[width=0.7\textwidth]{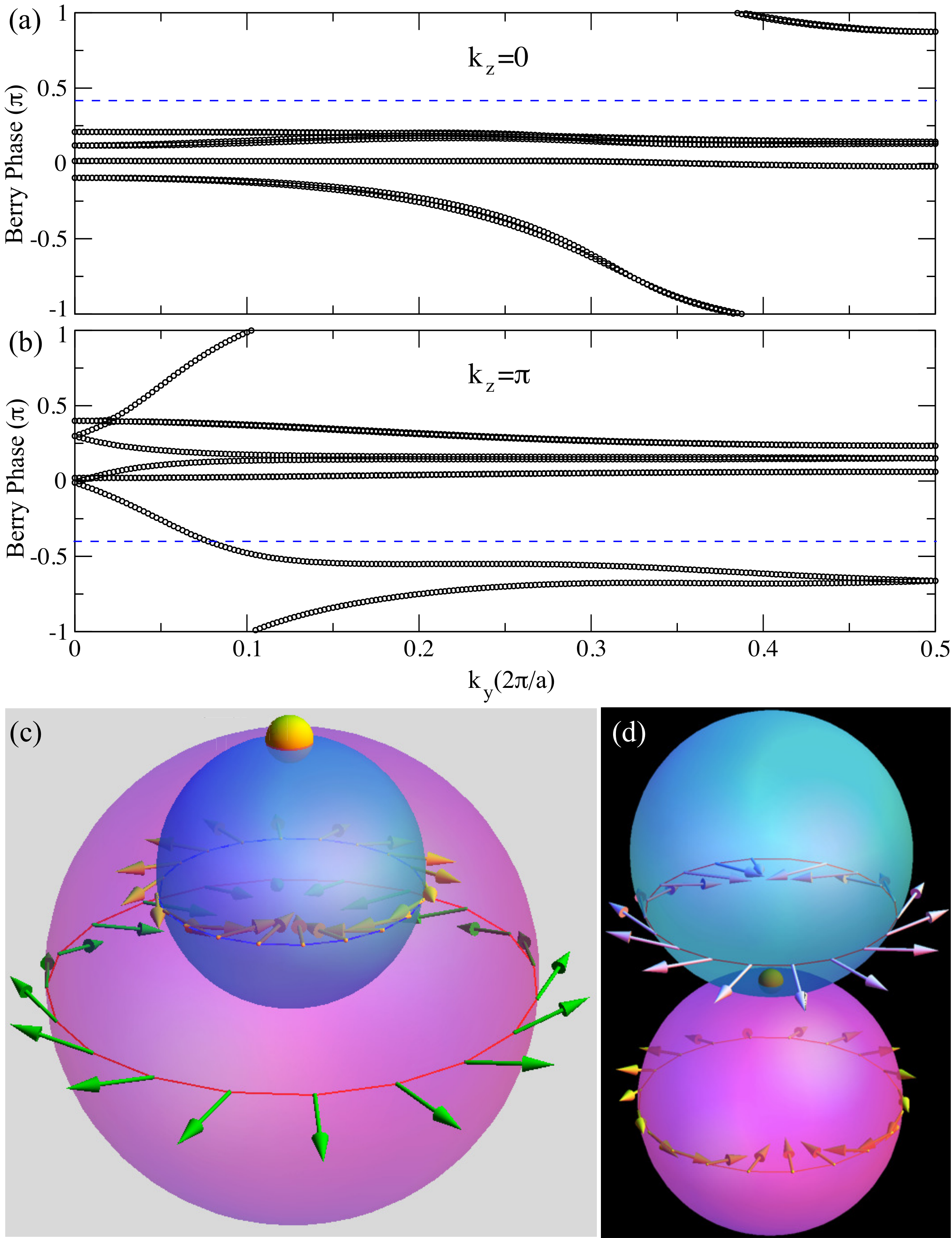}
\caption{(Color online) The eigenvalues of the Wilson loops along $k_x$-axis at fixed $k_y$ in $k_z$=0 (a) 
and $k_z$=$\pi$ (b) plane. The $k_z$=$\pi$ plane has nontrivial Z$_2$ number of 1. (c) and (d) are the schematic plot of
two tangent Fermi surface spheres enclosing two TDNPs with chemical potential sitting below (two hole pockets) and in-between (one electron and one hole pocket)
the TDNPs, respectively. The spin winding number of 2 on each sphere is also shown.
}
\label{TaN_wilson_loop}
\end{figure}

\begin{figure}
\includegraphics[width=0.7\textwidth]{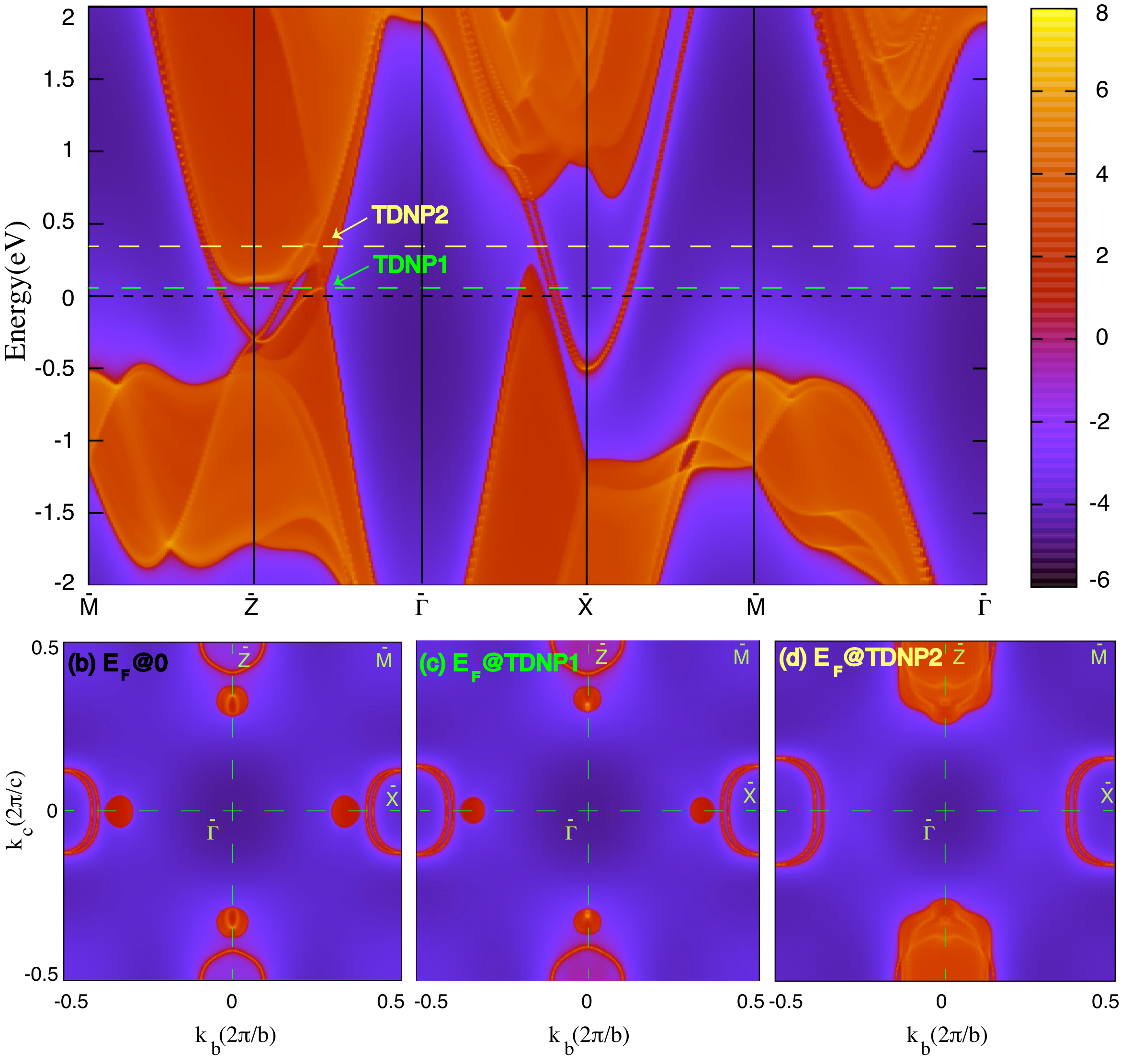}
\caption{
(Color online) TaN (100) surface state. (a) Band structure with weight projected onto one unit cell of top surface. (b), (c) and (d) are
Fermi surface with chemical potential at 0 eV, TDNP1 and TDNP2, respectively.
}
\label{TaN_100_Surface}
\end{figure}

Though $\theta$-TaN shows TDNPs along $\Gamma$-A and has both electron and hole Fermi pockets, its electronic structure
within the $k_z$=$\pi$ and $k_z$=0 plane can be looked as 2D insulators having time-reversal symmetry, which can give well defined $Z_2$ topological invariant to identify the band topology. Since $\theta$-TaN has no inversion symmetry, the Wilson loop method~\cite{YuRui_Z2_2011PRB, MRS_weng:9383312, adv_phys} is used to calculate this invariant. As shown in Fig.~\ref{TaN_wilson_loop}, the 2D electron bands in $k_z$=0 plane is trivial with $Z_2$ invariant being 0, while those in $k_z$=$\pi$ plane have $Z_2$=1. These two planes will have edge along
$\bar{\Gamma}$-$\bar{X}$ and $\bar{Z}$-$\bar{M}$, respectively, when cutting a plane [100] perpendicular to reciprocal
lattice vectors $\mathbf{b_1}$ or $\mathbf{b_2}$ in Fig.~\ref{crystructure}(b). Due to the different $Z_2$ number in $k_z$=0 and $k_z$=$\pi$
plane, the number of crossings between edge states and any in-gap energy level should be even and odd along 
$\bar{\Gamma}$-$\bar{X}$ and $\bar{Z}$-$\bar{M}$, respectively, shown in Fig.~\ref{TaN_100_Surface}. 
There is a Dirac cone like surface state centering $\bar{Z}$. The upper branch and lower branch connect to the conduction and valence
bands, respectively, in both $\bar{Z}$-$\bar{M}$ and $\bar{Z}$-$\bar{\Gamma}$ direction. The surface state around $\bar{X}$ are trivial
and both two branches connect conduction states. As we have shown, the TDNPs are protected by 
both $C_3$-axis and vertical mirror plane, which are broken on a side surface such as the [100]-surface, we hence do not expect features 
that are characteristic of the TDNP. 

Now we consider the spin structure near the a TDNP. A TDNP in our system is a crossing between a $J_z=3/2$ non-degenerate band and a $|J_z|=1/2$ degenerate band, so near the crossing point, the dynamics of the electronic states are governed by the following three-band Hamiltonian
\begin{equation}\label{eq:3band}
H_3(\mathbf{q})=\left[\begin{matrix}
				u_{1/2}q_z & \lambda_1q^2_+ & \lambda_2q_+\\
				\lambda_1q_-^2 & u_{1/2}q_z & \lambda_2q_-\\
				\lambda_2q_- & \lambda_2q_+ & u_{3/2}q_z
				\end{matrix}\right]
\end{equation}
where $\mathbf{q}$ is the momentum relative to TDNP, $u_{1/2,3/2}$ are the velocities of the two bands along $z$-axis, $\lambda_{1,2}$ are real constants and $q_\pm\equiv{q}_x\pm{i}q_y$. As we have shown, for any chemical potential near a TDNP, there are two carrier pockets touching each other at some point along $\Gamma$-A, where the Fermi level crosses the degenerate $|J_z|=1/2$-band. Eq.(\ref{eq:3band}) implies that the degenerate band will split away from $\Gamma$-A, and the energy split is quadratic in $\mathbf{q}$. Eq.(\ref{eq:3band}) also reveals the spin structure of the Fermi surface: if we identify the $J_z=\pm1/2$-state with spin up/down state, we find that along any horizontal loop on the Fermi surface, the spin winds exactly two rounds about the $z$-axis, and that the two touching Fermi surfaces have opposite windings. This is a topologically robust feature of the Fermi surface near our TDNP. In Fig.~\ref{TaN_wilson_loop}, we plot the schematics of the Fermi surfaces for two chemical potentials near TDNP, where there are two hole pockets (Fig.~\ref{TaN_wilson_loop}(c)) or one electron pocket and one hole pocket (Fig.~\ref{TaN_wilson_loop}(d)); on each Fermi surface, we plot the typical spin structure along some latitude.

{\it $k\cdot p$ model for $\theta$-TaN.} 

As discussed above, the low energy physics around the TDNPs and Fermi level are mostly determined by the bands
spanned by generate $J_{z}=\pm\frac{1}{2}$ and non-degenerate $J_{z}=\pm\frac{3}{2}$ states. A $k\cdot p$ effective model can be constructed
with these six states as basis set. The momentum zero point is set at A. 

Based on the orbital composition shown in Fig.~\ref{TaN_band} (a), the most relevant orbitals are the following $d$-orbitals of Ta: $d_{z^2}$, $d_{x^2-y^2}$ and $d_{xy}$. These orbitals plus spin degrees of freedom form the basis of the effective model: $\Psi=(d_{z^2\uparrow},id_{z^2\downarrow},id_{-2\downarrow},d_{+2\uparrow},id_{+2\downarrow},d_{-2\uparrow})^T$, where $d_{\pm2}\equiv{d}_{x^2-y^2}\pm{i}d_{xy}$. The derivation and the parameter fitting of the effective model is deterred to the Supplemental Materials, and in the main text, we briefly sketch the steps in its construction. First, we determine the little group of point A and how these basis states transform under the little group symmetries. Then we use the symmetry constraint
\begin{equation}
SH(\mathbf{q})S^{-1}=H(S\mathbf{q})
\end{equation}
to determine the form of $H(\mathbf{q})$ to a given order in $\mathbf{q}$, where $S$ is the matrix representation of a little group symmetry, $\mathbf{q}\equiv\mathbf{k}-A$ is the momentum relative to A, and $S\mathbf{q}$ is $\mathbf{q}$ acted by $S$. Finally, we use the dispersion from GGA to fit the parameters in the effective model.

%In Fig.~\ref{kpfit}, we show that the thus obtained $k\cdot{p}$ model can reasonably fit the dispersion along all high-symmetry lines near A.
%\begin{figure}
%\includegraphics[width=0.8\textwidth]{Figkp_fit_bands.eps}
%\caption{(Color online) The band structure from first-principles calculation (black line) can be well reproduced by 
%fitted k.p model (circles). (a) along A-$\Gamma$ (b) along A-L and (c) A-H. The Fermi surface in $k_x$-$k_z$ plane 
%with $k_y$=0 and 3D view with chemical potential at +70 meV obtained from fitted k.p model.
%}
%\label{kpfit}
%\end{figure}

\begin{figure}
\includegraphics[width=0.7\textwidth]{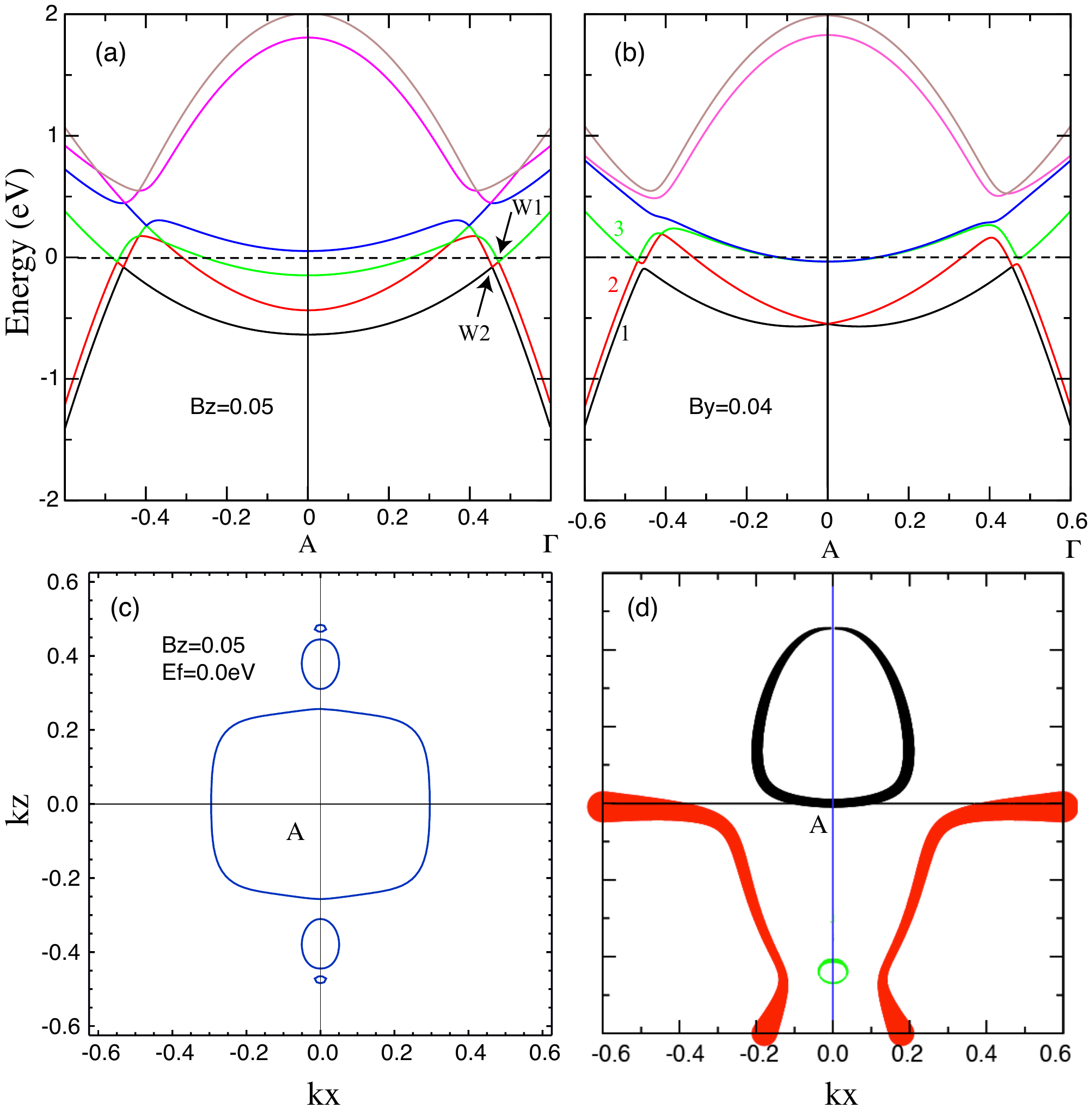}
\caption{(Color online) Band structure along A-$\Gamma$. (a) Magnetic field is applied along $z$ direction. The Weyl nodes W1 and W2 with opposite chirality are indicated. (b) Magnetic field is applied along $y$ direction. The bands associated with TDNP are labeled as band 1, 2 and 3. (c) The Fermi surface with chemical potential at 0 eV in (a). (d) The node-lines formed by band 1 and 2 (black and red ones) and band 2 and 3 (green). The thickness of the node-line represents the distance of nodal points to the Fermi level.
}
\label{magneticfield}
\end{figure}

An effective model helps us predict the effect of external fields. A uniform magnetic field induces a Zeeman field that couples to the spin degrees of freedom. The Zeeman field breaks time-reversal symmetry, but may preserve some crystal symmetry if applied along certain high-symmetry directions. For example, if $\mathbf{B}\parallel{\hat{z}}$, $C_3$ and $M_z$ symmetries are preserved while $T$ and $M_y$ broken. The TDNP point splits into two Weyl points with opposite monopole charges [see Fig.~\ref{magneticfield}(a)]; if $\mathbf{B}\parallel{\hat{y}}$, $M_y$ is preserved, while $C_3$ and $M_{y,z}$ are broken, and the TDNP splits into a nodal ring [see Fig.~\ref{magneticfield}(b)].

%For example, under a uniform Zeeman field, induced either by magnetic field or ferromagnetic doping, takes the form
%\begin{eqnarray}
%H_Z=\sum_{\tau\tau'}\mathbf{B}\cdot(g_1d^\dag_{z^2\tau}\mathbf{\sigma}_{\tau\tau'}d_{z^2\tau'}+g_2d^\dag_{x^2-y^2\tau}\mathbf{\sigma}_{\tau\tau'}d_{x^2-y^2\tau'}+g_2d^\dag_{xy\tau}\mathbf{\sigma}_{\tau\tau'}d_{xy\tau'}),
%\end{eqnarray}
%where $g_{1,2}$ are the $g$-factors of the $d_{z^2}$-orbital and the $d_{x^2-y^2,xy}$-orbitals. The Zeeman field breaks time-reversal symmetry, yet may preserve some crystal symmetry if applied along certain high-symmetry directions. For example, if $\mathbf{B}\parallel{\hat{z}}$, $C_3$ and $M_z$ symmetries are preserved while $T$ and $M_y$ broken. The three-band crossing point splits into two Weyl points with opposite monopole charges [see Fig.~\ref{magneticfield}(a)]; if $\mathbf{B}\parallel{\hat{y}}$, $M_y$ is preserved, while $C_3$ and $M_{y,z}$ are broken, and the three-band crossing splits into a nodal ring [see Fig.~\ref{magneticfield}(b)].

Another external field we consider is the strain tensor, which may be induced by curving the substrate or by applying a local force field using an atomic force microscope. The strain tensor is parameterized by five components, namely, $\epsilon_{x^2-y^2,xy,xz,yz,z^2}$, among which $\epsilon_{xz,yz}$ do not couple into our model. In the rest three components, $\epsilon_{z^2}$ do not break any symmetry, $\epsilon_{x^2-y^2}$ breaks $C_3$ and $\epsilon_{xy}$ breaks both $C_3$ and $M_y$ symmetries. Therefore, either $\epsilon_{x^2-y^2}$ or $\epsilon_{xy}$ will split the three-band crossing point, into line nodes and point nodes, respectively. (In Supplemental Materials, we explicitly write down the forms in which the Zeeman field and the strain tensor couple to the spin-orbital basis states.)

When a perturbation is added, the degenerate band of $|J_z|=1/2$ will split, so the touching Fermi surfaces will also separate. The separate Fermi surfaces may or may not have a Chern number, depending on the nature of the symmetry breaking perturbation. Also, we notice, in these examples, that independent of the form of perturbation, the three bands involved at a TDNP cannot be fully separated from each other in $k$-space by these perturbations: there still remain nodal lines or Weyl points between these bands. This is similar to the robustness of a Weyl point against all types of perturbations.

%and they couple to the three $d$-orbitals in the following way
%\begin{eqnarray}
%H_S=\sum_\tau\epsilon_{z^2}(\lambda_1d^\dag_{z^2\tau}d_{z^2\tau}+\lambda_2d^\dag_{+2\tau}d_{+2\tau}+\lambda_2d^\dag_{-2\tau}d_{-2\tau})+\epsilon_{+2}(\lambda_3d^\dag_{z^2\tau}d_{+2\tau}+\lambda_3d^\dag_{-2\tau}d_{z^2\tau}+\lambda_4d^\dag_{+2\tau}d_{-2\tau})+h.c.,
%\end{eqnarray}
%where $\epsilon_{\pm2}=\epsilon_{x^2-y^2}\pm{i}\epsilon_{xy}$ and we have assumed that all strain effects are onsite and have no spin dependence. Mark that under these assumptions, $\epsilon_{xz,yz}$ do not couple into our model. Among the rest three components, $\epsilon_{z^2}$ do not break any symmetry, $\epsilon_{x^2-y^2}$ breaks $C_3$ and $\epsilon_{xy}$ breaks both $C_3$ and $M_y$ symmetries. Therefore, either $\epsilon_{x^2-y^2}$ or $\epsilon_{xy}$ will split the three-band crossing point, into line nodes and point nodes, respectively.

\textit{Landau levels and helical anomaly}

The large, anisotropic negative magnetoresistance observed in Weyl semimetals is considered an indirect proof for chiral anomaly, a hallmark of the Weyl fermions. This anomaly means that the total electric current is not conserved ($\partial_\mu{J}^\mu\neq0$) on the quantum level, while the classical action remains invariant under the charge U(1) transform. The easiest way to see this is to consider the 3D Landau levels of Weyl fermions under a weak magnetic field along $z$-axis: the zeroth Landau levels at different $k_z$ constitute one chiral mode going along positive or negative $z$-axis, depending on the monopole charge of the Weyl point. Therefore the total number of modes going along $+z$-axis and $-z$-axis become different, that is, the total current is non-vanishing. To check if our new semimetals have any type of quantum anomaly, we start with looking at the 3D Landau levels under weak field.

On each side of A, there are two TDNP close to each other, where two non-degenerate $3/2$-bands cross one degenerate $1/2$-band. The states near these triple crossing points can hence be described by a four-band model. In the Supplemental Materials, we computed the Landau levels to the linear order of $k_{x,y}$ and field strength $B$ and we find that there are always two counter-propagating modes along $k_z$. Since the rotation axis is unbroken if the field is along $z$-axis, one can still label these modes by their respective $C_3$ eigenvalues, finding that one mode has $C_3=e^{-{i}\pi/3}$ and the other mode $C_3=e^{{i}\pi/3}$, where $\pm$ depends on whether the field is along $+z$-axis or $-z$-axis. Since the two modes have different $C_3$-eigenvalues, their crossing is symmetry-protected. In this case, the total charge current is zero as the two are counter-propagating, but net spin current is nonzero, because the two modes carry different $J_z$ (or $C_3$ eigenvalues). These two zero modes can be compared to the pair of helical edge modes of a quantum spin Hall (QSH) state in several aspects. First, crossing of the two modes is protected (i.e., cannot be gapped) by rotation symmetry in our case, and by time-reversal symmetry in the case of QSH state. Second, in both cases the two modes carry a net spin current. Finally, the back scattering between these modes are prohibited by rotation symmetry in our system and by time-reversal symmetry in QSH state. Such similarities suggest the name ``helical zeroth Landau level'' (HZLL) for the two modes. The existence of HZLL under small field indicates a new type of anomaly, termed `helical anomaly', that can be associated with this new type of semimetals. In terms of field theory, this anomaly means that while the classical action of effective theory for the Hamiltonian near the two TDNP on one side of A is invariant under $C_3$ rotation, the quantum partition function is not. Again, similar to the chiral anomaly, when the whole BZ is taken into account, the anomaly vanishes, as one can see from the fact that the two triple crossings on the other side of A contribute a pair of helical modes carrying an opposite spin current. Therefore, the helical anomaly is physically relevant only as long as the inter-valley scattering between the two sides of A is negligible.

\section{Discussion}

The TDNP in $\theta$-TaN should be compared with the recently proposed TDNP protected by nonsymmorphic symmetries. In the latter case, the TDNP appears at a high-symmetry point (BZ corner) and is pinned to that point, while in our case it is at a high-symmetry line and can move along the line by parameter tuning; in Ref.[\onlinecite{newfermion}], at an ideal integer filling, the Fermi surface shrinks to a point, and by breaking some crystalline symmetry, the system can be fully gapped, while in our case the Fermi surface always has a finite size; in Ref.[\onlinecite{newfermion}], the nonsymmorphic symmetries play a central role in protection of TDNP, while in $\theta$-TaN the space group is symmorphic and the TDNP is protected by rotation and mirror symmetries.

Compared to other topological semimetals, the topology of the FS and its evolution under the external fields are the key features of the ``New Fermion" state in $\theta$-TaN. With the increase of the chemical potential, the FS evolve from hole-hole type to electron-hole type and finally to electron-electron type. Two Lifshitz transitions happen accordingly when the chemical potential hits two TDNP. The existence of singular points on FS will lead to interesting phenomena in transport, for example in the quantum oscillation behavior under a magnetic field. For each separate piece of FS, the quantum oscillation under a weak field can be explained nicely by the semi-classical theory with the phase of the quantum oscillation being determined fully by the accumulation of the Berry phase along the extremal orbits. For systems having FS with touching points, the semi-classical orbits become undefined even at the low field due to the tunneling between two pieces of FS, a phenomenon known as the ``œmagnetic breakdown'' in quantum oscillation. Since the presence of the touching points on FS is protected by the $C_3$ rotation symmetry, $\theta$-TaN provides an ideal platform for the quantum transport studies for such systems.

\section{Conclusions}

In conclusion, the ``new fermions'' state with triply degenerate nodal points can be realized in $\theta$-TaN. The appearance of the nodal point is protected by the rotation symmetry and mirror symmetry, which allows both 2D and 1D representations along the $\Gamma$-A direction. Breaking these symmetries by the external fields will lead to either Weyl semimetal or nodal line semimetal phases. The Landau level calculation manifests the presence of helical anomaly in $\theta$-TaN. Finally, for an arbitrary Fermi level, our ``new fermion'' state hosts FSs that touch each other, leading to interesting transport properties, e.g., the possible magnetic breakdown in the quantum oscillation experiments. 

\section{Acknowledgments}
We acknowledge the supports from National Natural Science Foundation of China (Grant Nos. 11274359 and 11422428),
the National 973 program of China (Grant No. 2013CB921700) and the ``Strategic Priority Research Program (B)''
of the Chinese Academy of Sciences (Grant No. XDB07020100). Partial of the calculations were preformed on TianHe-1(A), the National Supercomputer Center in Tianjin, China.

\bibliography{TaN_ref}
\newpage
{\bf \large Supplemental Materials}
\section{Derivation of the effective model near A}
In this section, we derive the explicit form of the six-band effective model for the states near A, following the three steps sketched in the main text.

The little group at $A$ is generated by the following symmetry operations: threefold rotation $C_3$, mirror reflection about the $xy$-plane $M_z$, about the $xz$-plane $M_y$ and time-reversal $T$. They are represented by the following matrices in the chosen basis:
\begin{eqnarray}
C_3&=&\mathrm{diag}\{e^{-i\pi/3},e^{i\pi/3},e^{-i\pi/3},e^{i\pi/3},-1,-1\},\\
\nonumber
M_z&=&i\mathrm{diag}\{1,-1,-1\}\otimes\sigma_z,\\
\nonumber
M_y&=&iI_{3\times{3}}\otimes\sigma_x,\\
\nonumber
T&=&iKI_{3\times{3}}\sigma_y,
\end{eqnarray}
where $\sigma_{x,y,z}$ are the Pauli matrices. The effective Hamiltonian takes the bilinear form
\begin{eqnarray}
\hat{H}=\sum_\mathbf{q}\Psi^\dag(\mathbf{q})H(\mathbf{q})\Psi(\mathbf{q}),
\end{eqnarray}
where $\mathbf{q}\equiv\mathbf{k}-A$ is the momentum relative to A.
These symmetries constrain on the generic form of $H(\mathbf{q})$
\begin{eqnarray}
C_3^{-1}H(q_+,q_-,q_z)C_3&=&H(e^{i2\pi/3}q_+,e^{-i2\pi/3}q_-,q_z),\\
\nonumber
M_zH(q_+,q_-,q_z)M_z^{-1}&=&H(q_+,q_-,-q_z),\\
\nonumber
M_xH(q_+,q_-,q_z)M_x^{-1}&=&H(q_-,q_+,q_z),\\
\nonumber
TH(q_+,q_-,q_z)T^{-1}&=&H(-q_+,-q_-,-q_z).
\end{eqnarray}
The lowest order $k.p$-model obtained after applying the constraints takes the form
\begin{equation}
H(q_x,q_y,q_z)=\left[\begin{array}{cccccc}
a_{11}+b_{11} & C_{1}k_{+}^{2}q_{z} & a_{12}+b_{12} & C_{12}q_{-} & C_{13}q_zq_{+} & D_{13}q_{+}\\
\ast & a_{11}-b_{11} & C_{12}k_{+} & a_{12}-b_{12} & D_{13}q_{-} & C_{13}q_zq_{-}\\
\ast & \ast & a_{22}+b_{22} & C_{2}q_{+}^{2}q_{z} & iC_{23}q_{+} & iD_{23}q_{+}q_{z}\\
\ast & \ast & \ast & a_{22}-b_{22} & iD_{23}q_{-}q_{z} & iC_{23}q_{-}\\
\ast & \ast & \ast & \ast & a_{33}+b_{33} & C_{3}q_{z}\\
\ast & \ast & \ast & \ast & \ast & a_{33}-b_{33}
\end{array}\right],\label{eq:effH}
\end{equation}
where
\begin{eqnarray}\label{eq:parameters}
a_{11}&=&E_{1}+\frac{q_{x}^{2}+q_{y}^{2}}{2m_{xy1}}+\frac{q_{z}^{2}}{2m_{z1}},\\
\nonumber
a_{12}&=&iA_{12}q_z,\\
\nonumber
a_{22}&=&E_{2}+\frac{q_{x}^{2}+q_{y}^{2}}{2m_{xy2}}+\frac{q_{z}^{2}}{2m_{z2}},\\
\nonumber
a_{33}&=&E_3+D_{12}k_z(q_+^3-q_-^3),\\
\nonumber
b_{11}&=&iD_{1}(q_{+}^{3}-q_{-}^{3}),\\
\nonumber
b_{12}&=&D_{12}q_z(q_+^3-q_-^3),\\
\nonumber
b_{22}&=&iD_{2}(q_{+}^{3}-q_{-}^{3}),\\
\nonumber
b_{33}&=&-iD_3(q_+^3-q_-^3).
\end{eqnarray}

All parameters in Eq.(\ref{eq:parameters}) can be determined by fitting the dispersion of $H(\mathbf{q})$ in Eq.(\ref{eq:effH}) to the result from first principles calculation. They are found to be $E_1=1.9$, $m_{xy1}=0.23$, $m_{z1}=-0.056$, $A_{12}=0.38$, $E_2=-0.048$, $m_{xy2}=0.067$, $m_{z2}=0.21$, $E_3=-0.53$, $D_{12}=0.40$, $D_2=-0.85$, all in unit of eV.

\section{Coupling Zeeman field and strain tensor to the system}
Since the Zeeman field only couples to the spin but not the orbital degrees of freedom, and different orbitals may have different coupling strength, the lowest order coupling takes the form
\begin{eqnarray}
H_Z=\sum_{\tau\tau'}\mathbf{B}\cdot(g_1d^\dag_{z^2\tau}\mathbf{\sigma}_{\tau\tau'}d_{z^2\tau'}+g_2d^\dag_{x^2-y^2\tau}\mathbf{\sigma}_{\tau\tau'}d_{x^2-y^2\tau'}+g_2d^\dag_{xy\tau}\mathbf{\sigma}_{\tau\tau'}d_{xy\tau'}),
\end{eqnarray}
where $g_{1,2}$ are the $g$-factors of the $d_{z^2}$-orbital and the $d_{x^2-y^2,xy}$-orbitals.

On the other hand, the strain tensor is assumed to couple only to the orbital but not the spin degrees of freedom, so that the lowest order term reads
\begin{eqnarray}
H_S=\sum_\tau\epsilon_{z^2}(\lambda_1d^\dag_{z^2\tau}d_{z^2\tau}+\lambda_2d^\dag_{+2\tau}d_{+2\tau}+\lambda_2d^\dag_{-2\tau}d_{-2\tau})+\epsilon_{+2}(\lambda_3d^\dag_{z^2\tau}d_{+2\tau}+\lambda_3d^\dag_{-2\tau}d_{z^2\tau}+\lambda_4d^\dag_{+2\tau}d_{-2\tau})+h.c.,
\end{eqnarray}
where $\epsilon_{\pm2}=\epsilon_{x^2-y^2}\pm{i}\epsilon_{xy}$.

\section{Landau levels on one side of A}

Here we consider the effect of a weak field along $z$-axis on the band structure on one side of A, focusing on the two TDNP. From Fig.\~2(d), we see that a doubly degenerate band with $J_z=\pm{1/2}$ crosses the two non-degenerate $J_z=3/2$ bands, so the minimal model is a four-band model. If we use the basis $(|+1/2\rangle,|-1/2\rangle,|3/2_a\rangle,|3/2_b\rangle)^T$, along $\Gamma$-A, the Hamiltonian is diagonal
\begin{equation}
H_z(q_z)=diag\{\epsilon_{1/2}(q_z),\epsilon_{1/2}(q_z),\epsilon_{3/2,a}(q_z),\epsilon_{3/2,b}(q_z)\}.
\end{equation}
Since we are interested in the lowest several Landau levels under a weak magnetic field, we only keep the linear orders in $q_{x,y}$
\begin{equation}
H_\parallel=\left[\begin{matrix}
      0 & 0 & v_1q_+ & v_2q_+ \\
      * & 0 & v_2q_- & v_1q_-\\
      * & * & 0 & 0\\
      * & * & * & 0
   \end{matrix}\right].
\end{equation}
where $v_{1,2}$ are parameters implicitly depending on $q_z$. When a field is added along $z$-axis, we have the substitution
\begin{eqnarray}
q_+&\rightarrow&\sqrt{B}a^\dag\\
\nonumber
q_-&\rightarrow&\sqrt{B}a.
\end{eqnarray}
We convert to the following single fermion basis
\begin{equation}
(|+1/2,n\rangle,|-1/2,n-2\rangle,|3/2_a,n-1\rangle,|3/2_b,n-1\rangle)^T
\end{equation}
in which the Hamiltonian reads
\begin{equation}
H=\left[\begin{matrix}
	\epsilon_{1/2} & 0 & v_1\sqrt{B}\sqrt{n} & v_2\sqrt{B}\sqrt{n}\\
	0 & \epsilon_{1/2} & v_2\sqrt{B}\sqrt{n-1} & v_1\sqrt{B}\sqrt{n-1}\\
	v_1\sqrt{B}\sqrt{n} & v_2\sqrt{B}\sqrt{n-1} & \epsilon_{3/2,a} & 0\\
	v_2\sqrt{B}\sqrt{n} & v_1\sqrt{B}\sqrt{n-1} & 0 & \epsilon_{3/2,b}
	\end{matrix}\right]
\end{equation}
for $n\ge2$ and
\begin{eqnarray}
H_0&=&\epsilon_{1/2}\label{eq:n=0},\\
H_1&=&\left[\begin{matrix}
	\epsilon_{1/2} & v_1\sqrt{B} & v_2\sqrt{B}\\
	v_1\sqrt{B} & \epsilon_{3/2,a} & 0\\
	v_2\sqrt{B} & 0 & \epsilon_{3/2,b}
	\end{matrix}\right]\label{eq:n=1}.
\end{eqnarray}
for $n=0$ and $n=1$.
It is easy to see that for $n\ge2$, the spectrum is gapped, so the two ``zero'' modes come from $n=0$ and $n=1$, respectively. In fact, for $n=0$, Eq.(\ref{eq:n=1}) tells us that $|E_0\rangle=|+1/2,0\rangle$ is already an eigenstate with the energy $E_0=\epsilon_{1/2}$; for $n=1$, Eq.(\ref{eq:n=1}) shows that the other zero mode is a linear combination of three states, namely, $|-1/2,0\rangle$, $|3/2_a,1\rangle$ and $|3/2_b,1\rangle$. The explicit expression of the eigenstate or the eigenvalue is complicated, but for the special case of $v_1=0$ ($v_2=0$), we have the simple expressions $E_1=\epsilon_{3/2,a}$ ($E_1=\epsilon_{3/2,b}$) and $|E_1\rangle=|3/2_a,1\rangle$ ($|E_1\rangle=|3/2_b,1\rangle$). The two zero modes $E_0$ and $E_1$ in general will cross each other, but since $|E_0\rangle$ has $J_z=1/2$ and $|E_1\rangle$ has $J_z=-1/2$, the crossing point is protected by $C_3$-rotation symmetry.

\end{document}